\documentclass[10pt,twocolumn,letterpaper]{article}
\usepackage[pagenumbers]{cvpr}

\usepackage{enumitem}
\setlist{nosep, leftmargin=14pt}

\usepackage{graphicx}
\usepackage{color}
\usepackage{amsmath,amssymb,amsfonts}
\usepackage{algorithm}
\usepackage{algpseudocode}
\usepackage{mathtools}
\usepackage{multirow}
\usepackage[table,xcdraw,dvipsnames]{xcolor}
\usepackage{adjustbox}
\usepackage{subcaption}
\usepackage{booktabs}
\usepackage{mathtools}
\usepackage{listings}
\usepackage[font=scriptsize]{caption}
\usepackage{enumitem}
\usepackage{svg}

\algnewcommand{\Inputs}[1]{%
  \State \textbf{Inputs:}
  \Statex \hspace*{\algorithmicindent}\parbox[t]{.8\linewidth}{\raggedright #1}
}
\algnewcommand{\Initialize}[1]{%
  \State \textbf{Initialize:}
  \Statex \hspace*{\algorithmicindent}\parbox[t]{.8\linewidth}{\raggedright #1}
}

\usepackage[pagebackref,breaklinks,colorlinks,citecolor=cvprblue]{hyperref}
\usepackage[capitalize]{cleveref}
\crefname{section}{Sec.}{Secs.}
\Crefname{section}{Section}{Sections}
\Crefname{table}{Table}{Tables}
\crefname{table}{Tab.}{Tabs.}
\usepackage[dvipsnames]{xcolor}

\definecolor{cvprblue}{rgb}{0.21,0.49,0.74}

\def\paperID{*****} 
\def\confName{CVPR}
\def\confYear{2024}

\title{Motion-informed Needle Segmentation in Ultrasound Images}
\author{Raghavv Goel \thanks{Correspondence to raghavvg@alumni.cmu.edu.}
\quad
 Cecilia Morales 
 \quad
 Manpreet Singh
 \quad
 Artur Dubrawski
 \quad
 John Galeotti
 \quad
 Howie Choset \\
 Carnegie Mellon University \\
 }

\begin{document}


\maketitle

\begin{abstract}
Segmenting a moving needle in ultrasound images is challenging due to the presence of artifacts, noise, and needle occlusion. This task becomes even more demanding in scenarios where data availability is limited. 
In this paper, we present a novel approach for needle segmentation for 2D ultrasound that combines classical Kalman Filter (KF) techniques with data-driven learning, incorporating both needle features and needle motion. Our method offers three key contributions. First, we propose a compatible framework that seamlessly integrates into commonly used encoder-decoder style architectures. Second, we demonstrate superior performance compared to recent state-of-the-art needle segmentation  models using our novel convolutional neural network (CNN) based KF-inspired block, achieving a 15\% reduction in pixel-wise needle tip error and an 8\% reduction in length error. Third, to our knowledge we are the first to implement a learnable filter to incorporate non-linear needle motion for improving needle segmentation. 
\end{abstract}
\vspace{-0.2in}
\section{Introduction} \label{introduction}
\begin{figure}
    \centering
\includegraphics[width=0.40\textwidth,height=3cm]{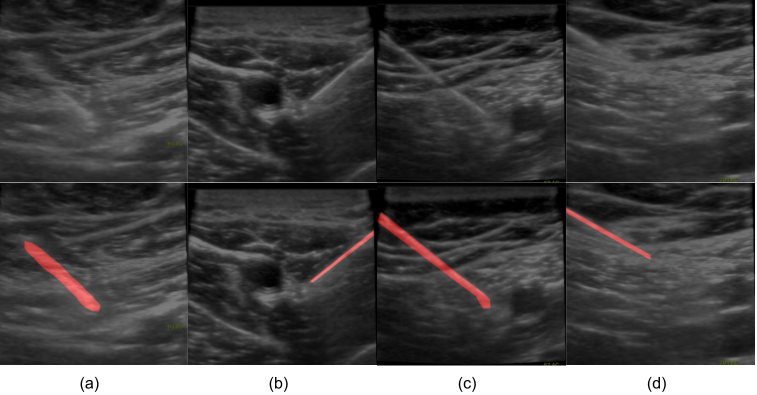}
    \caption{ Sample images from our ultrasound dataset. The top row shows the raw images and the bottom row shows the same images with the ground truth needle position overlaid in red. The images show (a) unclear initial half of needle, (b) multiple artifacts that look like the needle, (c)  needle occluded in the middle, and (d) needle-tip has poor visibility }
    \label{fig:sample_ultrasound_images}
\vspace{-0.1in}
\end{figure}

Needle insertion is commonly used in clinical diagnosis and treatment, providing a minimally invasive method for accessing a variety of medical conditions\cite{ChapmanGA}. Providing precise needle location during percutaneous ultrasound-guided needle insertion procedures can provide valuable insights to the clinician and improve the procedures' success \cite{gumbs2022advances}.
However, segmenting needles in ultrasound images is a challenging task. Ultrasound images are prone to having artifacts, noise, and shadows, which lead to a low signal-to-noise ratio \cite{1588392}. Moreover, the thin structure of the needle can be easily occluded by nerves or become partially visible due to bending out of the ultrasound image plane. Some of the aforementioned challenges are shown in Fig. \ref{fig:sample_ultrasound_images}.


%

Common deep learning based medical image segmentation methods have an encoder and a decoder. The encoder transforms the input image, capturing the image features, while the decoder further transforms and upsamples these features, outputting the corresponding prediction mask. This architecture can be seen in U-Net \cite{ronneberger2015u}, U-Net with Attention gates \cite{oktay2018attention}, TransUnet \cite{chen2021transunet}, Excitation Network \cite{excitation_network}, among others.
While these methods work well for segmenting static and complex anatomical structures, their efficacy in segmenting moving needles within ultrasound images has been found lacking as no needle motion information is used, ie. excitation networks are used to generate static needle masks. We believe that needle motion serves as a robust learning signal for needle segmentation. \color{black} 

Recent deep-learning methods for needle segmentation have also exploited relative change between consecutive image frames by using a stack of the current image together with prior frames as input to the encoder \cite{mukhopadhyay2021deep}, pixel-wise difference of consecutive image frames as an additional input to the encoder  \cite{mwikirize2019learning}, or using two different encoders for consecutive images leading to a single decoder \cite{chen2022automatic, dunnhofer2020siam}, or adding separate upsampling blocks for needle-tip and needle-shaft \cite{gao2021robust}.
However, these simple operations for capturing needle movement perform poorly when the background is non-stationary. Some methods \cite{mwikirize2019single,8091815} track only the needle-tip, 
however, these method are useful only in cases when needle-tip movement and brightness are more distinct than the rest of the image, as is the case when the needle is inserted orthogonal to the ultrasound image view, which is not consistent with our data.    

Algorithms for needle tracking involving KF \cite{geraldes2014neural, yan2021needle} make impractical assumptions by naively modeling needle dynamics as linear with needle moving at a constant velocity. In general, during skin penetration, the needle follows a complex motion with non-linear dynamics as it interacts with tissues, organs, and arteries which exert forces at the needle tip and along the shaft which can result in needle bending. Hence, needle motion can't be derived analytically.    

\begin{figure}[h]
\centering
\includegraphics[width=0.5\textwidth]{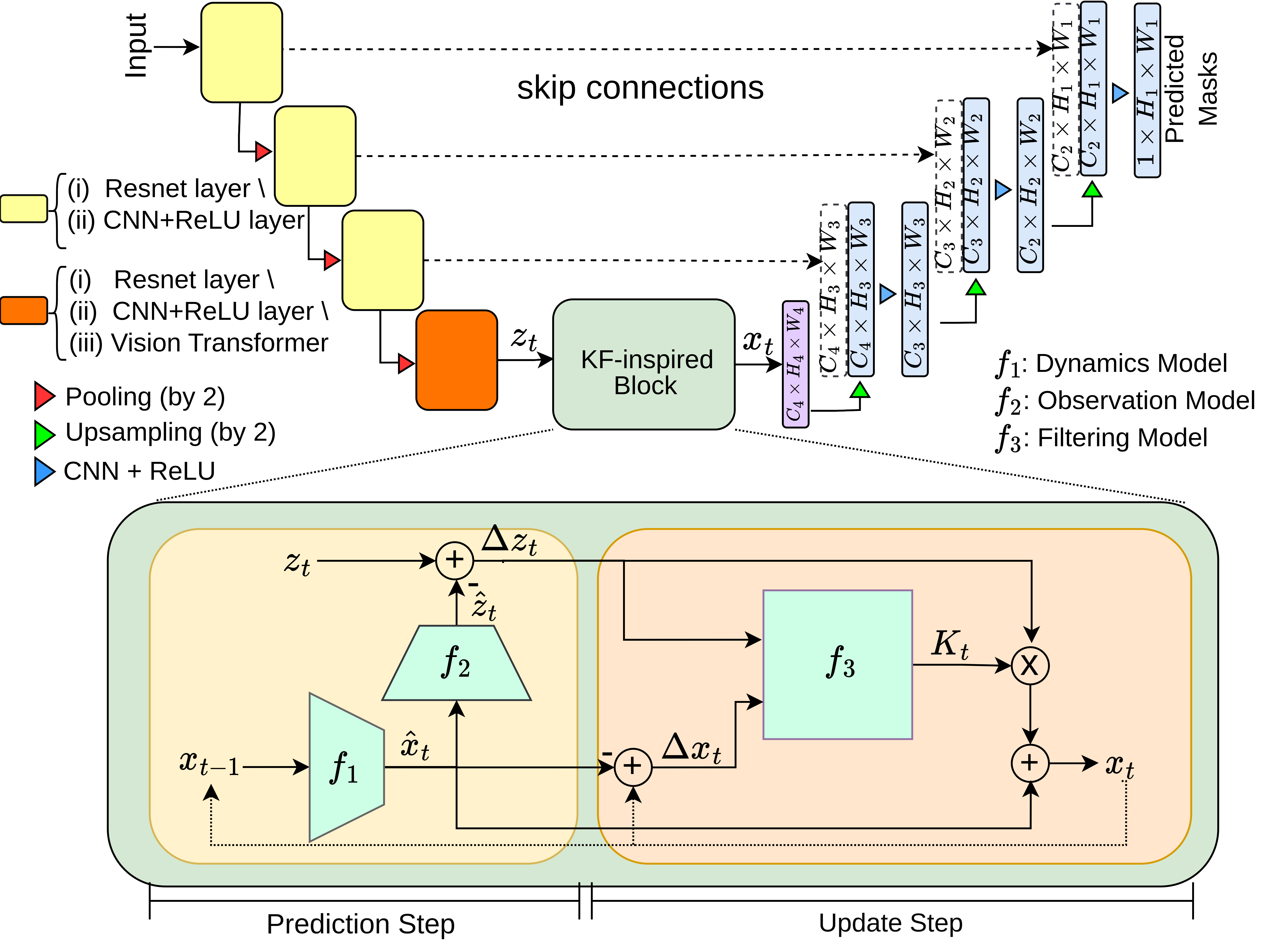}
\caption{
Overall architecture diagram showing the variety of encoders in yellow and orange. The yellow blocks can be either (i) residual network blocks or, (ii) convolutional neural network (CNN) and ReLU blocks. The following orange block can be either (i), (ii), or (iii) a vision transformer block. 
The red triangles are the max-pooling layer to downsample the features. The right side in light blue shows decoder with convolutional and ReLU layers (blue triangle) and upsampling (green triangle). Finally, our KF-inspired block (light green) is in between the encoder and decoder having two stages: prediction step and update step, with three learnable blocks represented by $f_1, f_2, f_3$.
}
\label{fig:ConvKalman_Unet}
\end{figure}

We therefore propose a KF-inspired block (Fig. \ref{fig:ConvKalman_Unet}) which learns the needle motion features as a dynamics model in the feature space of the input images and, then combines these motion features with the visual features for improved needle segmentation. Our block stands out in that the feature space of the input images has features corresponding to needle-like structures; however,artifacts are present, therefore it combines the needle motion by acting as a filter to remove undesirable artifacts. 

\label{sec:related_work}

\vspace{-0.2in}
\section{Method}
\label{sec:method}
\color{red} \color{black}
The overall method for segmenting needles in ultrasound images consists of three stages: (i) an \textbf{encoder} for extracting visual features of the needle from individual frames, (ii) a \textbf{KF-inspired block} for capturing needle motion across successive image frames and combining motion with visual features, and (iii) a \textbf{decoder} with skip-connections to combine high-level features with the feature representation infused with needle motion to produce segmentation masks. These three stages are depicted in Fig. \ref{fig:ConvKalman_Unet}. 

As shown in Fig. \ref{fig:ConvKalman_Unet}, the initial $3$ encoder blocks (yellow) \color{black} consist of either convolutional and ReLU layers, similarly to the encoder blocks in U-Net architectures \cite{ronneberger2015u}, or residual-network (ResNet) blocks \cite{he2016deep}. If the first three blocks are based on the encoder in U-Net, then the last block (orange) \color{black} can either be a block of convolutional and ReLU layer or a Vision transformer (ViT), resembling the encoder in TransUNet \cite{chen2021transunet} where the placement of ViT block follows  \cite{chen2021transunet}. In contrast, for ResNet blocks, the last block is also a residual network block, and we use ResNet-18 due to computing budget. For the decoder, we follow the decoder mentioned in \cite{chen2021transunet} which consists of deconvolutional, convolutional and ReLU layers.All implementation has been done in Python using PyTorch for ease of reproducibility.

\subsection{KF-inspired Block}
We propose a learnable KF-inspired block improving upon KalmanNet \cite{revach2022kalmannet}  
with two key modifications. First, using CNN layer instead of linear layer, and second, placing the KF-inspired block in between an encoder and a decoder for the purpose of segmentation. The CNN helps to maintain and use the spatial structure of the feature space as no reshaping of the features is required which could result in loss of spatial information, as is the case for linear layer. The KF-inspired block is placed after the encoder to learn needle motion using lower level needle-features.
\color{black}

As seen from Fig. \ref{fig:ConvKalman_Unet}, the KF-inspired block follows similar high-level steps to the conventional KF \cite{kalman_filter}: prediction-step and update-step. For a time-step $t$, first, the prediction-step consists of the following: the observation ($z_t$), which represents the current image features, the state-estimate ($\hat{x}_{t}$), obtained by propagating output state at the previous time-step ($x_{t-1}$) through the dynamics model ($f_1$), and observation estimate ($\hat{z}_{t}$), produced by propagating $\hat{x}_{t}$ through the observation model ($f_2$).  Subsequently, the update-step consists of the following: computing dynamics error ($\Delta x_{t}=x_{t-1}-\hat{x}_{t}$) and observation error ($\Delta z_t = z_{t}-\hat{z}_{t}$), computing the Kalman gain ($K_{t}$) using history of dynamics and observation error propagated through $f_3$.
The Kalman gain acts as a filter combining the observation, dynamics, and computing the output state ($x_{t} = \hat{x}_{t} + K_{t} \odot \Delta z_{t}$). 

Generally, the observation, represented by the current image feature, has no information of needle motion and can even consist of undesirable features due to low signal to noise ratio of ultrasound images.
We hypothesize that by incorporating both the needle motion and needle features, we can improve the needle segmentation mask prediction. The KF is useful in such situations which involve noisy observations and approximate dynamics. The KF combines them both and produces an improved state through the update-step.

\subsection{Dataset}
Our dataset consists of 17 ultrasound videos of the femoral area of different participants, which were captured using a Butterfly handheld ultrasound probe with patients undergoing adductor canal (AC) blocks at the Brooke Army Medical Center (BAMC). The videos were collected as part of routine clinical trials, with approval from the Institutional Review Board (IRB). During these clinical procedures, doctors often have to exert jerking motions with the needle in order to visualize it properly. It is worth noting that even though a rigid needle is used during the procedures, when interacting with the tissue the needle can sometimes buckle due to resistance from the surrounding tissues. The ultrasound probe 
had a maximum depth of $10$cm. Each video contains between $250$ to $600$ frames captured at a resolution of around $580 \times 585$ and a frame-rate of $20$ fps. Although images were resized to $256 \times 256$. 
The annotations were created by expert clinicians using the open-source Computer Vision Annotation Tool (CVAT) \cite{boris_sekachev_2020_4009388} to ensure accuracy. The annotations provide segmentation masks for various features including nerve structures and needle positions. It is important to note that the ultrasound data is obtained by a non-stationary ultrasound probe. 

\subsection{Comparison Models}
\label{subsec:comparison_model}

To evaluate the efficacy of our method in an encoder-decoder framework, we show the performance of our KF-inspired block with different encoders: (i) encoder from vanilla U-Net architecture \cite{ronneberger2015u}, (ii) ResNet-18, and (iii) encoder from TransUnet architecture \cite{chen2021transunet}. Using these encoders and decoder from \cite{chen2021transunet}, we make a variety of different architectures by replacing/removing the KF-inspired block: (a) no block, (b) no block with attention gates 
\footnote{https://github.com/ozan-oktay/Attention-Gated-Networks}, (c) no block but inputting a stack of consecutive images to the encoder instead of a single image, following \cite{mukhopadhyay2021deep} 
, (d) Long-Short Term Memory (LSTM)~\cite{6795963} unit block, (e) ConvLSTM block~\cite{10.5555/2969239.2969329}. Note that no block present ((a), (b) and (c)) implies that $x_t = z_t$ from Fig. \ref{fig:ConvKalman_Unet}. These different combinations can recreate known architectures, for example, the encoder (i) with (a) or (b) and the given decoder gives the vanilla U-Net or U-Net with gates Attention architecture respectively. Similarly, the encoder (iii) with (a) and given decoder gives the TransUnet from \cite{chen2021transunet}. We utilized LSTM and ConvLSTM blocks to demonstrate that using CNNs to capture motion directly in the feature space (utilizing adjacency information), as done in ConvLSTM and our method, outperforms capturing motion in a smaller embedding space, without using adjacency information, as done in LSTM and KalmanNet. This is particularly evident in the challenging task of needle segmentation, as shown in Table \ref{table:result_comparison} below. 
\color{black}

\subsection{Training Setup}
We utilized 5-fold cross-validation for both training and testing. Our dataset consisted of 17 videos, which were divided into folds based on the number of frames each video had. Each fold contained approximately an equal number of frames, with a range of 1200-1520 frames per fold. We randomized the splits at the video level, ensuring no frames from the same video were present in both training and testing sets to prevent data leakage. During preprocessing, frames were subjected to variable rotation and jitter for data augmentation. 
In each fold, the model was trained on four sets and tested on the remaining set, repeating this process five times to ensure robust results. 
The final performance metrics were derived by averaging the results and are presented in Table \ref{table:result_comparison}. Due to the limited size of the dataset, no validation set was made for fine-tuning. 

The encoder (i) from Subsection \ref{subsec:comparison_model} was initialized with Kaiming initialization \cite{he2015delving} and ResNet-18 was initialized using Imagenet weights \footnote{Vision-Transformer initialized from github.com/Beckschen/TransUNet
}.
We used Binary Cross-Entropy (BCE) loss\footnote{\href{https://pytorch.org/docs/stable/generated/torch.nn.BCEWithLogitsLoss.html}{Pytorch's BCELossWithLogits} is used.}, and AdamW \cite{loshchilov2017decoupled} optimizer with default hyper-parameters, with a learning rate of $0.001$, and \textit{ReduceLROnPlateau} learning-rate scheduler with a factor of $0.7$ and patience of $20$ for training the models end-to-end. 
We used a batch size of $4$ and $8$ for the models with ((d),(e),(f) in Subsection \ref{subsec:comparison_model}) and without ((a),(b),(c) in Subsection \ref{subsec:comparison_model}) a block in between the encoder-decoder respectively. We couldn't use a higher batch size for the case with blocks due to limited hardware resources and back-propagation through time (BPTT) requiring more memory. A batch for models with a block ((d),(e),(f) in Subsection \ref{subsec:comparison_model}) has a sequence of consecutive frames (7-10) where each frame is passed into the model one at a time and then after the last frame is passed, the model weights are updated through BPTT to capture the needle motion. \color{black} 
On average, each model is trained for 10 epochs which comprises more than 6000 optimization steps in total. All the models are trained using a 24 GB Nvidia RTX 3090Ti. 
During inference, a single ultrasound image is propagated through the network. The frameworks having a block between the encoder-decoder maintain a hidden state which captures the dependence of previously propagated frames and is used to provide additional context to the current frame.
\subsection{Metrics}
We use common image segmentation metrics: Dice Score-Coefficient (DSC), Precision, and Recall along with needle-specific metrics such as needle tip error ($\Delta x, \Delta y$) and needle length error ($\Delta L$). The additional metrics are useful for segmenting needle-like objects which occupy a small number of pixels as compared to total pixels in the image. Needle tip error is calculated based on the absolute difference between last pixel along the needle shaft of ground truth mask and predicted mask. Similarly, needle length can be determined by calculating the $L_2$ norm between the coordinates of the first and last pixel along the needle shaft, then we take absolute difference of length between the ground truth and predicted mask to compute the needle length error. The reported errors are averaged over the all the frames in the test data. 
\vspace{-0.15in}
\section{Results}
\label{sec:experiment_result}
\begin{figure}[t]
    \centering
    \includegraphics[width=1\columnwidth]{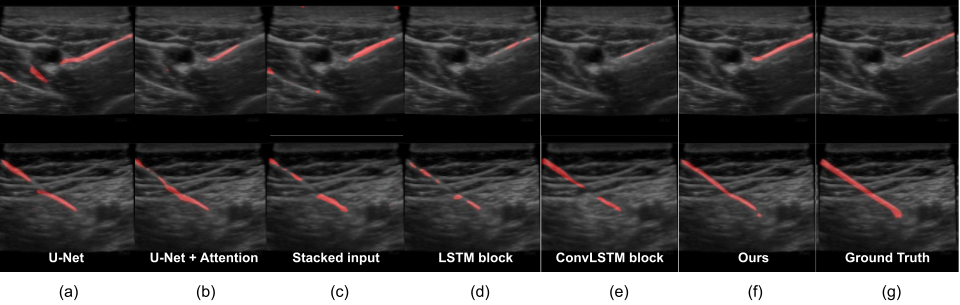}
    \caption{Qualitative performance with encoder from vanilla U-Net architecture and decoder from \cite{chen2021transunet} which results in (a) U-Net, (b) U-Net with Attention gates, (c) stacked input, (d) LSTM block, (e) ConvLSTM block, (f) \textbf{Ours}, and (g) ground truth label. }
    \label{fig:ours_vs_all_qualitative}
\end{figure}

\begin{table}[h]
\caption{Comparison of dice score-coefficient (DSC), precision, recall, needle tip error ($\Delta x, \Delta y$) and needle length error ($\Delta L$) with different encoders averaged over $5$ folds. The overall best are \textbf{bolded} while the best in each encoder are in \color{blue} blue \color{black}. Note: An upward-pointing arrow ($\uparrow$) indicates better performance at higher values, while a downward-pointing arrow ($\downarrow$) indicates better performance at lower values. Naming convention for encoders is V = encoder from vanilla U-Net architecture, R = ResNet-18, and H = encoder from TransUnet architecture. In addition to the encoders, there are variations in the model's architecture (a) no block, (b) no block with attention gates, (c) no block with inputting a stack of consecutive 5 images, (d) Long-Short term Memory unit block, (e) ConvLSTM block}
\centering
\resizebox{\columnwidth}{!}{
\begin{tabular}{l|c | c |c |c |c |c} 
 \hline
  Method  & DSC(D) 	$\uparrow$ & Precision(P) 	$\uparrow$ & Recall(R)    	$\uparrow$ & $\Delta x$  	$\downarrow$ & $\Delta y$ 	$\downarrow$ & $\Delta L$ 	$\downarrow$\\ 
 \hline 
      V
           & $0.32 \pm 0.15$ 
           & $0.37 \pm 0.19$ 
           & $0.33 \pm 0.14$ 
           & $54 \pm 18$ 
           & $91 \pm 29$ 
           & $110 \pm 43$
    \\
      V$+$(b)
            & $0.32 \pm 0.14$
            & $0.38 \pm 0.17$
            & $0.34 \pm 0.15$
            & $48  \pm 20$
            & $74 \pm 35$
            & $76 \pm 38$
     \\
      V$+$(c) 
            & $0.34 \pm 0.13$
            & $0.42 \pm 0.18$
            & $0.33 \pm 0.11$
            & $48 \pm 18$
            & $83 \pm 35$
            & $110 \pm 50$
     \\
     V$+$(d)  
            & $0.22 \pm 0.10$
            & $0.34 \pm 0.11$
            & $0.28 \pm 0.10$
            & $59 \pm 9$
            & $80 \pm 20$
            & $87 \pm 28$
     \\
      V$+$(e) 
            & \color{blue} \textbf{0.39} $\pm$ \textbf{0.17}
            & 0.44 $\pm$ 0.16
            &\color{blue} \textbf{0.43} $\pm$ \textbf{0.19}
            & $43 \pm 7$
            & $67 \pm 17$
            & $80 \pm 27$            
     \\
      V$+$\textbf{Ours}
            &\color{blue} \textbf{0.39} $\pm$ \textbf{0.16}
            &\color{blue} \textbf{0.48} $\pm$ \textbf{0.19}
            & $0.39 \pm 0.15$
            & \color{blue} 35 $\pm$ 12
            &\color{blue} 55 $\pm$ 26
            & \color{blue} 72 $\pm$ 32   
     \\
\hline 

        R
            & $0.27 \pm 0.14$
            & $0.39 \pm 0.22$
            & $0.27 \pm 0.10$
            & $53 \pm 17$
            & $73 \pm 21$
            & $74 \pm 14$     
     \\
      R$+$(b)  
            & $0.29 \pm 0.13$
            & $0.41 \pm 0.18$
            & $0.27 \pm 0.09$
            & $57 \pm 8$
            & $79 \pm 20$
            & $87 \pm 18$     
     \\
      R$+$(c) 
            & $0.28 \pm 0.15$
            & $0.41 \pm 0.20$
            & $0.26 \pm 0.12$
            & $43 \pm 12$
            & \color{blue} 65 $\pm$ 29
            & $73 \pm 23$     
     \\
      R$+$(d)  
            & $0.24 \pm 0.13$
            & $0.40 \pm 0.14$
            & $0.30 \pm 0.16$
            & $64 \pm 23$
            & $72 \pm 21$
            & $89 \pm 23$     
     \\
      R$+$(e) 
            & $0.34 \pm 0.15$
            & $0.44 \pm 0.18$
            & $0.34 \pm 0.18$
            & $49 \pm 17$
            & $73 \pm 28$
            & $78 \pm 28$     
     \\
      R$+$\textbf{Ours} 
            &\color{blue} $0.36 \pm 0.16$
            &\color{blue} $0.45 \pm 0.20$
            &\color{blue} $0.36 \pm 0.19$
            &\color{blue} $40 \pm 17$
            &\color{blue} $65 \pm 32$
            &\color{blue} $71 \pm 27$     
     \\
 \hline 
    
     H
            & $0.29 \pm 0.13$
            & $0.41 \pm 0.21$
            & $0.27 \pm 0.11$
            & $40 \pm 12$
            & $63 \pm 28$
            & $72 \pm 19$       
     \\
      H$+$(b)  
            & $0.29 \pm 0.13$
            & $0.39 \pm 0.17$
            & $0.29 \pm 0.10$
            & $48 \pm 19$
            & $65 \pm 32$
            & $69 \pm 28$          
     \\
      H$+$(c) 
            & $0.30 \pm 0.15$
            & $0.41 \pm 0.20$
            & $0.30 \pm 0.14$
            & $37 \pm 14$
            & $64 \pm 30$
            & $67 \pm 16$          
     \\
      H$+$(d)  
            & $0.27 \pm 0.12$
            & $0.40 \pm 0.15$
            & \color{blue} $0.38 \pm 0.14$
            & $60 \pm 23$
            & $72 \pm 19$
            & $91 \pm 17$          
     \\
      H$+$(e) 
            &\color{blue} $0.36 \pm 0.16$
            &\color{blue} 0.47 $\pm$ 0.20
            & $0.34 \pm 0.17$
            & $38 \pm 9$
            & $56 \pm 26$
            &\color{blue} \textbf{60} $\pm$ \textbf{32}          
     \\
      H$+$\textbf{Ours} 
            & \color{blue} $0.36 \pm 0.16$
            & \color{blue} $0.47 \pm 0.20$
            & $0.35 \pm 0.18$
            &\color{blue} \textbf{34} $\pm$ \textbf{8}
            &\color{blue} \textbf{53} $\pm$ \textbf{24}
            & 66 $\pm$ 23    
     \\
 \hline
\end{tabular}
}
\label{table:result_comparison}
\end{table}

We employed both qualitative and quantitative methods to evaluate the effectiveness of our approach in needle tracking. The performance of the three different encoders and the six different choices with and without a block in between the encoder and decoder is compared in
 Table \ref{table:result_comparison}. For the sake of brevity in the results table, we refer to the encoders as \textbf{V} for the encoder from vanilla U-Net architecture, \textbf{R} for ResNet-18, and \textbf{H} for encoder from TransUnet architecture.
 
 Our results show that our method consistently achieves the best or second-best results across all encoders, showing that our proposed block is generalizable, trainable, and compatible with different encoder-decoder networks, and can give visible performance improvement when no KF-inspired block is used. Specifically, our method outperforms all other methods in terms of precision and needle tip error, while also achieving the highest dice score together with ConvLSTM. Knowing the precise location of the needle tip in needle-guided procedures is critical to ensuring their success. 
 The needle tip error metric plays a vital role in determining whether the needle has accurately reached the intended vessel. In this regard, the use of the KF-inspired block has proven to be highly effective, as it yields a smaller needle tip error compared to the diameter of the femoral vessels. This increased level of accuracy can greatly improve the success rates of procedures, leading to better patient outcomes and overall safety.

The encoder from vanilla U-Net architecture shows overall best results on DSC, Precision, and Recall scores with ConvLSTM and KF-inspired blocks. However, for the ConvLSTM and KF-inspired blocks, the encoder from TransUnet architecture performs slightly better overall in terms of needle length and tip error. In general, training the vanilla encoder from scratch has proven to be advantageous, likely due to the specific characteristics of our dataset. The use of pre-trained ImageNet weights as a warm-start for the ResNet and Vision-transformer has not been as effective, potentially due to (1) low signal-to-noise ratio in the ultrasound images, and (2) our ultrasound dataset differs significantly from the high-resolution colored images used for pre-training.
 
\begin{figure}[h!]
    \centering
    \includegraphics[width=0.8\columnwidth, height=2.75cm]{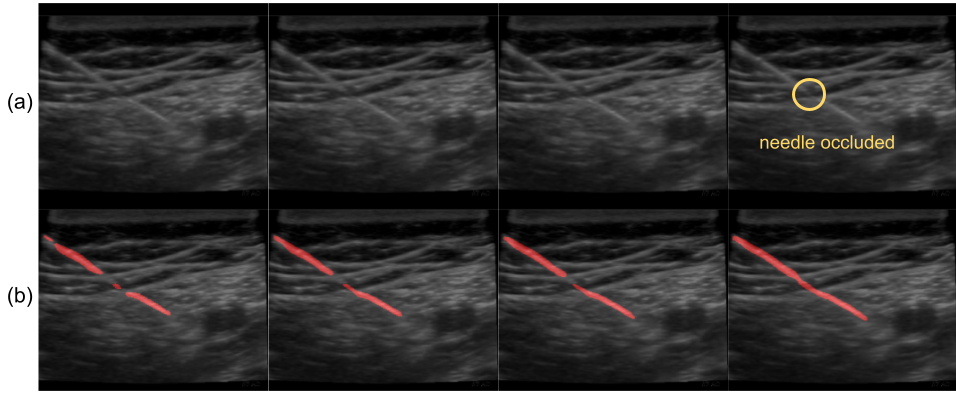}
    \caption{Qualitative assessment of the proposed method on test data. (a) Sequential Ultrasound Images: The figure showcases a series of ultrasound images capturing a needle occluded within nervous tissue. This occlusion is positioned at the center of the images. (b) Needle Mask Prediction Overlay: Overlaid in red upon the ultrasound images is the needle mask prediction generated by our proposed method. Notably, the predictive accuracy of the mask consistently improves over time, underscoring the efficacy of our approach. This observed behavior parallels that of a Kalman Filter (KF), where predictions tend to refine as time progresses.}
    \label{fig:conv_kalman_sequence}
\end{figure}

 Qualitative results on a sample of ultrasound images from the test set are shown in Fig. \ref{fig:ours_vs_all_qualitative}, the proposed network's needle mask resembles the ground truth the most and correctly estimates the location of the needle. 
 Furthermore, from Fig.\ref{fig:conv_kalman_sequence} we observe compelling evidence of the effectiveness and progressive refinement inherent in our proposed method's prediction especially in the presence of occlusion showing similarity to a KF, due to its filtering structure. 

\vspace{-0.25in}
\section{Conclusion}
\label{sec:conclusion}

This research introduces a novel KF-inspired block that enhances the accuracy of needle segmentation mask prediction by integrating both needle features and motion. This adaptable and learnable block is compatible with different encoders. 
Our experiments demonstrate that our approach achieves superior results in terms of DSC, Precision, and needle tip error, ranking second in performance of other metrics. Furthermore, our method outperforms or competes favorably with others within each encoder type. 
Finally, our approach opens up avenues for further research, such as applying our method to different video domains, finding the optimal placement of the KF-inspired block, comparing the affect of different encoder blocks such as Swin transformer \cite{liu2021swin}, combining other needle segmentation methods with our KF-inspired block, incorporating ideas from learnable optical flow. 
{
    \small
    \bibliographystyle{ieeenat_fullname}
    \bibliography{main}
}

%





\end{document}